\documentclass[twocolumn,preprintnumbers,prl,nofootinbib,superscriptaddress]{revtex4-1}

\usepackage{amsmath,amssymb,slashed}
\usepackage{hyperref}
\usepackage{url}
\usepackage{breakurl}
\usepackage{graphicx}
\usepackage{mathtools,xparse}

\def\bra#1{\langle #1 |}
\def\ket#1{| #1 \rangle}

\def\app#1#2{%
  \mathrel{%
    \setbox0=\hbox{$#1\sim$}%
    \setbox2=\hbox{%
      \rlap{\hbox{$#1\propto$}}%
      \lower1.1\ht0\box0%
    }%
    \raise0.25\ht2\box2%
  }%
}

\DeclarePairedDelimiter{\norm}{\lVert}{\rVert}

\begin{document}

\title{Spacetime Equals Entanglement}

\author{Yasunori Nomura}
\affiliation{Berkeley Center for Theoretical Physics, Department of Physics, 
  University of California, Berkeley, CA 94720, USA}
\affiliation{Theoretical Physics Group, Lawrence Berkeley National 
  Laboratory, Berkeley, CA 94720, USA}
\affiliation{Kavli Institute for the Physics and Mathematics 
 of the Universe (WPI), The University of Tokyo Institutes for Advanced Study, 
 Kashiwa 277-8583, Japan}
\author{Nico Salzetta}
\affiliation{Berkeley Center for Theoretical Physics, Department of Physics, 
  University of California, Berkeley, CA 94720, USA}
\affiliation{Theoretical Physics Group, Lawrence Berkeley National 
  Laboratory, Berkeley, CA 94720, USA}
\author{Fabio Sanches}
\affiliation{Berkeley Center for Theoretical Physics, Department of Physics, 
  University of California, Berkeley, CA 94720, USA}
\affiliation{Theoretical Physics Group, Lawrence Berkeley National 
  Laboratory, Berkeley, CA 94720, USA}
\author{Sean J. Weinberg}
\affiliation{Berkeley Center for Theoretical Physics, Department of Physics, 
  University of California, Berkeley, CA 94720, USA}
\affiliation{Theoretical Physics Group, Lawrence Berkeley National 
  Laboratory, Berkeley, CA 94720, USA}

\begin{abstract}
We study the Hilbert space structure of classical spacetimes under 
the assumption that entanglement in holographic theories determines 
semiclassical geometry.  We show that this simple assumption has profound 
implications; for example, a superposition of classical spacetimes may 
lead to another classical spacetime.  Despite its unconventional nature, 
this picture admits the standard interpretation of superpositions of 
well-defined semiclassical spacetimes in the limit that the number of 
holographic degrees of freedom becomes large.  We illustrate these ideas 
using a model for the holographic theory of cosmological spacetimes.
\end{abstract}

\maketitle

\section{Introduction}

How does the semiclassical picture arise from the fundamental theory 
of quantum gravity?  Recently it has become increasingly clear that 
quantum entanglement in holographic~\cite{'tHooft:1993gx,Susskind:1994vu} 
descriptions plays an important role in the emergence of the classical 
spacetime of general relativity~\cite{Ryu:2006bv,Hubeny:2007xt,%
VanRaamsdonk:2009ar,Swingle:2012wq,Maldacena:2013xja,Sanches:2016sxy}. 
This raises the possibility that entanglement is indeed the defining 
property that controls the physics of dynamical spacetimes.

In this letter we take the view that entanglement in holographic 
theories {\it determines} gravitational spacetimes at the semiclassical 
level.  Rather than proving this statement, we adopt it as a guiding 
principle and explore its consequences.  This principle has profound 
implications for the structure of the Hilbert space of quantum gravity. 
In particular, it allows us to obtain a classical spacetime as a 
superposition of (an exponentially large number of) different classical 
spacetimes.  We show that despite its unconventional nature, this 
picture admits the standard interpretation of superpositions of 
well-defined semiclassical spacetimes in the limit that the number 
of holographic degrees of freedom becomes large.

To illustrate these concepts, we use a putative holographic theory 
for cosmological spacetimes, in which the effects appear cleanly. 
Our basic points, however, persist more generally; in particular, 
we expect that they apply to a region of the bulk in the 
AdS/CFT correspondence~\cite{Maldacena:1997re}.  In the context 
of Friedmann-Robertson-Walker (FRW) universes, we find an interesting 
``Russian doll'' structure:\ states representing a universe filled 
with a fluid having an equation of state parameter $w$ are obtained 
as exponentially many (exponentially rare) superpositions of those 
having an equation of state with $w' > w$ ($< w$).

While completing this work, we received Ref.~\cite{Almheiri:2016blp} 
by Almheiri, Dong and Swingle which studies how holographic entanglement 
entropies are related to linear operators in the AdS/CFT correspondence. 
Their analysis of the thermodynamic limit of the area operators 
overlaps with ours.  See also Ref.~\cite{Papadodimas:2015jra} 
for related discussion.

\section{Holographic Theory on Screens}
\label{sec:framework}

We begin by describing the holographic framework we work in.  The 
AdS/CFT case appears as a special situation of this more general 
(albeit more conjectural) framework.

The covariant entropy bound~\cite{Bousso:1999xy} implies that the 
entropy on a null hypersurface generated by a congruence of light rays 
terminated by a caustic or singularity is bounded by its largest cross 
sectional area ${\cal A}$ divided by $2$ in Planck units.  (The entropy 
on each side of the largest cross sectional surface is bounded by 
${\cal A}/4$.)  This suggests that for a fixed gravitational spacetime, 
the holographic theory lives on a hypersurface---called the holographic 
screen---on which null hypersurfaces foliating the spacetime have the 
largest cross sectional areas~\cite{Bousso:1999cb}.

The procedure of erecting a holographic screen has a large ambiguity. 
A particularly useful choice~\cite{Nomura:2011dt,Nomura:2011rb} is 
to adopt an ``observer centric reference frame.''  Let the origin of 
the reference frame follow a timelike curve $p(\tau)$ which passes 
through a fixed spacetime point $p_0$ at $\tau = 0$, and consider the 
congruence of past-directed light rays emanating from $p_0$.  Assuming 
the null energy condition, the light rays focus toward the past, and 
we may identify the apparent horizon, i.e.\ the codimension-2 surface 
on which the expansion of the light rays vanishes, to be an equal-time 
hypersurface---called a leaf---of a holographic screen.  Repeating 
the procedure for all $\tau$, we obtain a specific holographic screen, 
with the leaves parameterized by $\tau$, corresponding to foliating 
the spacetime region accessible to the observer at $p(\tau)$.  Such 
a foliation is consonant with complementarity~\cite{Susskind:1993if} 
which asserts that a complete description of a system refers only 
to the spacetime region that can be accessed by a single observer.

With this construction, we can view a quantum state of the holographic 
theory as living on a leaf of the holographic screen obtained as 
above.  We can then consider the collection of all possible quantum 
states on all possible leaves, obtained by considering all timelike 
curves in all spacetimes.  It is often convenient to consider 
Hilbert space ${\cal H}_B$ spanned by the states living on the 
``same'' leaf $B$.%
\footnote{In general, the equivalence condition for the label $B$ is 
 not well understood.  For states representing FRW universes, however, 
 we expect from the high symmetry of the system that $B$ is uniquely 
 specified by the leaf area (at least in some coarse sense).}
We can then write the full Hilbert space 
as~\cite{Nomura:2011dt,Nomura:2011rb}
\begin{equation}
  {\cal H} = \sum_B {\cal H}_B + {\cal H}_{\rm sing},
\label{eq:H}
\end{equation}
where ${\cal H}_{\rm sing}$ contains intrinsically quantum gravitational 
states that do not admit a spacetime interpretation, and we have defined 
the sum of Hilbert spaces by%
\footnote{Unlike Ref.~\cite{Nomura:2011rb}, here we do not assume 
 specific relations between ${\cal H}_B$'s or ${\cal H}_{\rm sing}$. 
 In particular, ${\cal H}_{B_1}$ and ${\cal H}_{B_2}$ for different 
 leaves $B_1$ and $B_2$ may not be orthogonal.}
\begin{equation}
  {\cal H}_1 + {\cal H}_2 
  = \{ v_1 + v_2\, |\, v_1 \in {\cal H}_1, v_2 \in {\cal H}_2 \}.
\label{eq:H_sum}
\end{equation}
This formulation is not restricted to descriptions based on fixed 
semiclassical spacetime backgrounds.  For example, we may consider 
a state in which macroscopically different universes are superposed.
Time evolution of a quantum gravity state occurs within the space 
of Eq.~(\ref{eq:H}).

Recently, Bousso and Engelhardt have identified two special classes 
of holographic screens~\cite{Bousso:2015mqa,Bousso:2015qqa}:\ if a 
portion of a holographic screen is foliated by marginally anti-trapped 
(trapped) surfaces, then that portion is called a past (future) 
holographic screen.  They proved that the area of leaves ${\cal A}(\tau)$ 
monotonically increases (decreases) for a past (future) holographic 
screen.  Furthermore, Ref.~\cite{Sanches:2016pga} proved that this 
area law holds locally.  In many regular circumstances, including 
expanding FRW universes, the holographic screen is a past holographic 
screen, so that the area of the leaves monotonically increases, 
$d{\cal A}(\tau)/d\tau > 0$.  In this letter we focus on this case.

What is the structure of the holographic theory and how can we explore 
it?  Recently, a conjecture has been made in Ref.~\cite{Sanches:2016sxy} 
which relates geometries of general spacetimes to the entanglement 
entropies of states in the holographic theory.  This extends the analogous 
statement~\cite{Ryu:2006bv,Hubeny:2007xt,Lewkowycz:2013nqa} in AdS/CFT 
to more general cases.  In particular, Ref.~\cite{Sanches:2016sxy} proved 
that for a given region $\Gamma$ of a leaf $\sigma$, a codimension-2 
extremal surface $E[\Gamma]$ anchored to the boundary $\partial \Gamma$ 
of $\Gamma$ is fully contained in the causal region $D_\sigma$ of 
$\sigma$:\ the domain of dependence of an interior achronal hypersurface 
whose only boundary is $\sigma$.  This implies that the normalized area 
of the extremal surface $E[\Gamma]$
\begin{equation}
  S[\Gamma] = \frac{1}{4} \norm{E[\Gamma]},
\label{eq:S_Sigma}
\end{equation}
satisfies expected properties of entanglement entropy, so that it can 
be identified with the entanglement entropy of the region $\Gamma$ in 
the holographic theory.  Here, $\norm{x}$ represents the area of $x$. 
If there are multiple extremal surfaces in $D_\sigma$ for a given 
$\Gamma$, then we must take the one with the minimal area.

\section{Holography for FRW Universes}
\label{sec:FRW}

Adopting the above framework, we now study the holographic description 
of $(3+1)$-dimensional FRW universes
\begin{equation}
  ds^2 = -dt^2 + a^2(t) \left[ \frac{dr^2}{1-\kappa r^2} 
    + r^2 (d\psi^2 + \sin^2\!\psi\, d\phi^2) \right],
\label{eq:FRW-metric}
\end{equation}
where $a(t)$ is the scale factor, and $\kappa < 0$, $= 0$ and $> 0$ for 
open, flat and closed universes, respectively.  We choose the origin 
of the reference frame, $p(\tau)$, to be at $r = 0$.  The holographic 
theory then lives on the holographic screen at
\begin{equation}
  r = \frac{1}{\sqrt{\dot{a}^2(t_*) + \kappa}} 
    \equiv r_\sigma(t_*),
\label{eq:r_AH}
\end{equation}
where the dot represents $t$ derivative, and $t_*$ is the FRW time on 
a leaf.  For flat and open universes, the leaves always form a past 
holographic screen as long as the universe is initially expanding. 
Below, we focus on these cases.

For the purpose of illustrating our points, it is sufficient to consider 
a ``single'' Hilbert space ${\cal H}_* \in \{ {\cal H}_B \}$ specified 
by a fixed leaf area ${\cal A}_*$.  Specifically, we consider FRW 
universes with $\kappa \leq 0$ having vacuum energy $\rho_\Lambda$ 
and filled with various ideal fluid components.  For every universe 
with $\rho_\Lambda < 3/2 {\cal A}_*$, there is an FRW time $t_*$ 
at which the area of the leaf is ${\cal A}_*$.  Any quantum state 
representing the system at such a time is an element of ${\cal H}_*$.

How does a state in ${\cal H}_*$ encode information about the universe 
it represents?  Consider an FRW universe with the energy density given 
by $\rho(t)$.  We can then determine the FRW time $t_*$ at which the 
leaf $\sigma_*$ has the area ${\cal A}_*$.  Now, consider a spherical 
cap region of $\sigma_*$ specified by an angle $\gamma$ ($0 \leq \gamma 
\leq \pi$):
\begin{equation}
  L(\gamma):\ \,\, 
  t = t_*, \quad r = r_\sigma(t_*), \quad 0 \leq \psi \leq \gamma,
\label{eq:L_gamma}
\end{equation}
and determine the extremal surface $E(\gamma)$ anchored on the boundary 
of $L(\gamma)$.  The quantity
\begin{equation}
  S(\gamma) = \frac{1}{4} \norm{E(\gamma)},
\label{eq:S_gamma}
\end{equation}
then gives the entanglement entropy of the region $L(\gamma)$ in the 
holographic theory.

We focus on the case in which the expansion of the universe is dominated 
by a single fluid component with $w$ or negative spacetime curvature 
in (most of) the region probed by the extremal surfaces anchored 
to $\sigma_*$.  This holds for almost all $t$ in realistic FRW 
universes.  In this case, $S(\gamma)$ becomes extensive with respect 
to ${\cal A}_*$~\cite{NSSW}, so that
\begin{equation}
  \tilde{S}(\gamma) = \frac{S(\gamma)}{{\cal A}_*/4},
\label{eq:tilde-S}
\end{equation}
is independent of ${\cal A}_*$.  This effectively counts the number 
of Bell pairs between $L(\gamma)$ and its complement per (qubit) degree 
of freedom.

We may express $\tilde{S}(\gamma)$ as a function of the fractional volume 
that $L(\gamma)$ occupies on $\sigma_*$
\begin{equation}
  F(\gamma) = \frac{1}{2} (1 - \cos\gamma),
\label{eq:tilde-V}
\end{equation}
i.e.\ $s(F) \equiv \tilde{S}\left(\gamma(F)\right)$.
\begin{figure}[t]
\begin{center}
  \includegraphics[height=4cm]{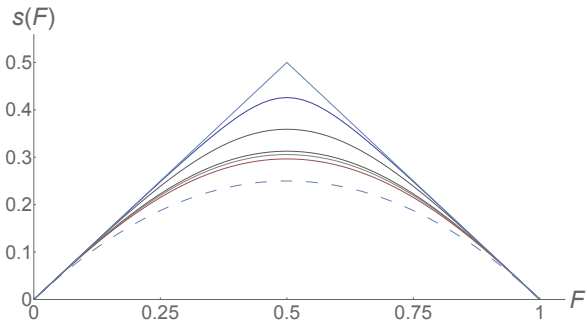}
\end{center}
\caption{The normalized entanglement entropy $s(F)$ for flat universes 
 with $w = -1$, $-0.98$, $-0.8$, $0$, $1/3$, $1$ (solid lines, from 
 the top to the bottom) and for a curvature dominated open universe 
 (dashed line).}
\label{fig:tilde-S}
\end{figure}
In Fig.~\ref{fig:tilde-S}, we plot this quantity for flat universes 
with $w = -1$ (vacuum energy), $-0.98$, $-0.8$, $0$ (matter), $1/3$ 
(radiation), $1$ and for a curvature dominated open universe.%
\footnote{In the case of a strictly single $w = -1$ component or 
 an exactly empty open universe, the leaf is located at an infinite 
 affine distance from $p_0$.  We view this as a mathematical 
 idealization.  Realistic universes are obtained, e.g., by introducing 
 an infinitesimally small amount of additional matter.}
This shows how a state in the holographic theory encodes the information 
about the spacetime it represents.  For example, $s(F)$ decreases 
monotonically in $w$ for any fixed $F$.

There are simple geometric bounds on $s(F)$.  The maximin 
construction~\cite{Sanches:2016sxy,Wall:2012uf} states that the 
extremal surface is the one having the maximal area among all possible 
codimension-2 surfaces each of which is anchored on $\partial L(\gamma)$ 
and has minimal area on some interior achronal hypersurface bounded 
by $\sigma_*$.  This implies that $s(F)$ obeys~\cite{NSSW}
\begin{equation}
  F(1-F) \leq s(F) \leq \frac{1}{2} 
    - \Bigl( \frac{1}{2}-F \Bigr)\, {\rm sgn}\Bigl( \frac{1}{2}-F \Bigr).
\label{eq:S-bounds}
\end{equation}
The universes dominated by a $w = -1$ component and curvature saturate 
the upper and lower limits, respectively.

\section{Qubit Model}
\label{sec:toy}

Unlike the case of asymptotically AdS spacetimes, entanglement entropies 
in the holographic theory for FRW universes obey a volume law.  (From 
the viewpoint of the holographic theory, ${\cal A}_*$ is a volume.) 
This motivates us to consider the following toy model for holographic 
states representing FRW universes.

Consider a Hilbert space for $N$ ($\gg 1$) qubits $\mathcal{H} = 
({\mathbf C}^{2})^{\otimes N}$.  Let $\Delta$ ($\leq N$) be a nonnegative 
integer and consider a typical superposition of $2^\Delta$ product states
\begin{equation}
  \ket{\psi} = \sum_{i=1}^{2^\Delta} 
    a_i\, \ket{x^i_1} \ket{x^i_2} \cdots \ket{x^i_N},
\label{eq:psi}
\end{equation}
where $\{ a_i \}$ is a normalized complex vector, and $x^i_{1,\cdots,N} 
\in \{ 0,1 \}$.  Given an integer $n$ with $1 \leq n < N$, we can break 
the Hilbert space into a subsystem $\Gamma$ for the first $n$ qubits 
and its complement $\bar{\Gamma}$.  We are interested in computing 
the entanglement entropy of $\Gamma$, $S_{\rm EE}(\Gamma)$.

Suppose $n \leq N/2$.  If $\Delta \geq n$, then $i$ in Eq.~(\ref{eq:psi}) 
runs over an index that takes many more values than the dimension of 
the Hilbert space for $\Gamma$, so that Page's argument~\cite{Page:1993df} 
tells us that $\Gamma$ has maximal entanglement entropy:\ $S_{\rm EE}(\Gamma) 
= n \ln 2$.  On the other hand, if $\Delta < n$ then the number of terms 
in Eq.~(\ref{eq:psi}) is much less than both the dimension of the Hilbert 
space of $\Gamma$ and that of $\bar{\Gamma}$, which limits the entanglement 
entropy:\ $S_{\rm EE}(\Gamma) = \Delta \ln 2$.

We therefore obtain
\begin{equation}
  S_{\rm EE}(\Gamma) = 
    \begin{cases}
      n      & n \leq \Delta, \\
      \Delta & n > \Delta,
    \end{cases}
\label{eq:S_EE-1}
\end{equation}
for $\Delta < N/2$, while
\begin{equation}
  S_{\rm EE}(\Gamma) = n,
\label{eq:S_EE-2}
\end{equation}
for $\Delta \geq N/2$.  Here and below, we drop the irrelevant factor 
of $\ln 2$.  The value of $S_{\rm EE}(\Gamma)$ for $n > N/2$ is obtained 
from $S_{\rm EE}(\Gamma) = S_{\rm EE}(\bar{\Gamma})$ since $\ket{\psi}$ 
is pure.

The behavior of $S_{\rm EE}(\Gamma)$ in 
Eqs.~(\ref{eq:S_EE-1},~\ref{eq:S_EE-2}) is reminiscent of that 
of $s(F)$ in Fig.~\ref{fig:tilde-S}.  The correspondence is 
given by
\begin{align}
  \frac{n}{N} &\,\leftrightarrow\, F,
\label{eq:corresp-1}\\
  \frac{\Delta}{N} &\,\leftrightarrow\, s\Bigl(\frac{1}{2}\Bigr),
\label{eq:corresp-2}
\end{align}
for $\Delta \leq N/2$.%
\footnote{States with $\Delta > N/2$ cannot be discriminated from those 
 with $\Delta = N/2$ using $S_{\rm EE}(\Gamma)$ alone.  The identity 
 of these states is not clear.  Below, we focus on the states with 
 $N/4 \leq \Delta \leq N/2$.}
In fact, we can consider the $N = {\cal A}_*/4$ qubits to be distributed 
over $\sigma_*$ with each qubit occupying a volume of $4$ in the 
holographic theory.  The identification of Eq.~(\ref{eq:corresp-1}) 
is then natural.  The quantity $\Delta$ controls what universe a state 
represents.  For fixed $\Delta$, different choices of the product 
states $\ket{x^i_1}\ket{x^i_2} \cdots \ket{x^i_N}$ and the coefficients 
$a_i$ give $e^N$ independent microstates for the FRW universe 
with $w = f(\Delta/N)$.  The function $f$ is determined by 
Eq.~(\ref{eq:corresp-2}); in particular, $f = -1$ ($> -1$) 
for $\Delta/N = 1/2$ ($< 1/2$).%
\footnote{For the present purpose, the curvature dominated universe 
 can be regarded as the universe filled with a fluid having $w = 
 +\infty$:\ $f = +\infty$ for $\Delta/N = 1/4$.}

\begin{figure}[t]
\begin{center}
  \includegraphics[height=4.5cm]{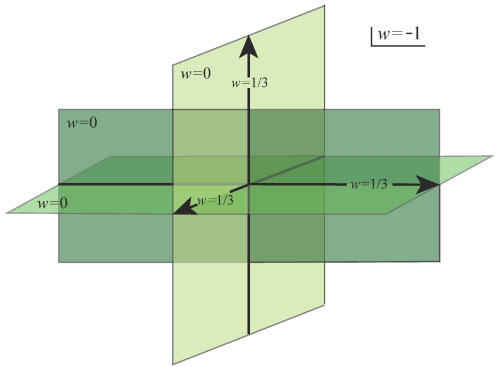}
\end{center}
\caption{A sketch of the Hilbert space for FRW universes, where we 
 have chosen three values of $w = -1,0,1/3$ for illustrative purposes. 
 Note that while the figure is drawn in 3-dimensional space, the 
 actual dimension of the Hilbert space (as well as that of a region 
 with a fixed $w$) is exponentially large.  In particular, to go 
 outside a region with a fixed $w$, an exponentially large number 
 of microstates must be superposed.  The same is true to go to 
 a smaller region with $w' > w$.}
\label{fig:sketch}
\end{figure}

Below, we assume that this model captures essential features of the 
holographic theory.%
\footnote{The holographic theory may have degrees of freedom representing 
 the region outside $D_{\sigma_*}$, which may be entangled with those 
 described here.  We assume that this does not affect the analyses 
 below based on $x^i_{1,\cdots,N}$ unless we probe more than half of 
 them; the extra degrees of freedom may become relevant if we consider 
 $n \geq N/2$.  This property indeed appears if the extra degrees of 
 freedom are modeled by an additional $N$ qubits, and the FRW states 
 are taken as typical superpositions of $2^\Delta$ product states in 
 the enlarged Hilbert space.}
An important point is that the set of states with a fixed $\Delta$ 
does not comprise a Hilbert space.  Moreover, the set of states with 
any fixed $\Delta$ spans the {\it entire} Hilbert space, containing 
all FRW universes corresponding to all values of $\Delta$.  For example, 
we may obtain a state with any $w' < w$ by superposing $e^{\Delta_{w'} 
- \Delta_w}$ states with $\Delta_w$, where $ \Delta_w \equiv N f^{-1}(w)$. 
We may also obtain a state with $w' > w$ as a superposition of carefully 
chosen $e^{\Delta_w}$ states with $\Delta_w$.%
\footnote{These superpositions must also change the matter content 
 filling the universe.}
These statements do not depend on the details of the model and are 
manifestations of the fact that entanglement, and thus spacetime 
geometry, cannot be represented by a linear operator at the microscopic 
level.  Figure~\ref{fig:sketch} depicts a sketch of the Hilbert space 
structure described here.

\section{Superpositions}
\label{sec:superposition}

One might worry that the Hilbert space structure described above 
may not allow for a consistent interpretation of superpositions of 
semiclassical universes.  For example, consider two states representing 
universes with $w_1$ and $w_2$ ($w_1 > w_2$), which contain respectively 
$e^{\Delta_1}$ and $e^{\Delta_2}$ product states ($\Delta_1 < \Delta_2$). 
Their superposition contains $e^{\Delta_1} + e^{\Delta_2} \approx 
e^{\Delta_2}$ product states.  Does this mean that a superposition 
of $w_1$ and $w_2$ universes always leads to a $w_2$ universe, making 
any reasonable many worlds interpretation of spacetime impossible?

Below we show that this is not the case.  In particular, information 
about each semiclassical spacetime is contained in the exponentially 
differing size of the coefficients when the state is expanded in the 
product state basis.

We regard universes with the equation of state parameters falling 
in a range $\delta w \ll 1$ to be macroscopically identical.  Here, 
$\delta w$ is a small number that does not scale with ${\cal A}_*$. 
We then find that a superposition of less than $e^{O(\delta w {\cal A}_*)}$ 
microstates representing a universe with some $w$ leads only to another 
microstate representing the same $w$ universe.  In other words, $e^N 
= e^{{\cal A}_*/4}$ microstates of the universe with {\it any} $w$ 
form an ``effective vector space'' unless we consider a superposition 
of an exponentially large, $\gtrsim e^{O(\delta w {\cal A}_*)}$, 
number of microstates.

How about a superposition of states representing universes with different 
$w$'s?  Consider two {\it normalized} microstates of the form given in 
Eq.~(\ref{eq:psi}):
\begin{align}
  \ket{\psi_1} &= \sum_{i = 1}^{e^{\Delta_1}} 
    a_i \ket{x_1^i x_2^i \cdots x_N^i},
\label{eq:psi_1}\\
  \ket{\psi_2} &= \sum_{i = 1}^{e^{\Delta_2}} 
    b_i \ket{y_1^i y_2^i \cdots y_N^i},
\label{eq:psi_2}
\end{align}
where $N/4 \leq \Delta_1 \neq \Delta_2 \leq N/2$, and the coefficients 
$a_i$ and $b_i$ are random as are the binary values $x^i_{1,\cdots,N}$ 
and $y^i_{1,\cdots,N}$.  We are interested in understanding the physical 
meaning of the normalized superposition
\begin{equation}
  \ket{\psi} = c_1 \ket{\psi_1} + c_2 \ket{\psi_2}.
\label{eq:psi-superp}
\end{equation}

The reduced density matrix for the first $n$ qubits ($n < N/2$) is
\begin{equation}
  \rho_n = {\rm tr}_{n+1 \cdots N} \ket{\psi} \bra{\psi}.
\label{eq:rho_n}
\end{equation}
Because of the normalization conditions
\begin{equation}
  \sum_{i=1}^{e^{\Delta_1}} |a_i|^2 
  = \sum_{i=1}^{e^{\Delta_2}} |b_i|^2 = 1,
\label{eq:norm}
\end{equation}
$\rho_n$ takes the form
\begin{equation}
  \rho_n = |c_1|^2 \rho_n^{(1)} + |c_2|^2 \rho_n^{(2)},
\label{eq:rho_n-sep}
\end{equation}
up to corrections exponentially suppressed in $N$.  (For a detailed 
derivation, see Ref.~\cite{NSSW}.)  Here, $\rho_n^{(1)}$ ($\rho_n^{(2)}$) 
are the reduced density matrices we would obtain if the state were 
genuinely $\ket{\psi_1}$ ($\ket{\psi_2}$):
\begin{equation}
  \rho_n^{(1)} = \sum_{i=1}^{e^{\Delta_1}} 
    |a_i|^2 \ket{x_1^i \cdots x_n^i} \bra{x_1^i \cdots x_n^i},
\label{eq:rho_n-1}
\end{equation}
with $\rho_n^{(2)}$ given by $\Delta_1 \rightarrow \Delta_2$, $a_i 
\rightarrow b_i$, and $x^i_{1,\cdots,n} \rightarrow y^i_{1,\cdots,n}$. 
The matrix $\rho_n$ thus takes the form of an incoherent 
classical mixture.  Moreover, the simple form of the matrices 
in Eqs.~(\ref{eq:rho_n-sep},~\ref{eq:rho_n-1}) implies that the 
entanglement entropies are also incoherently added
\begin{equation}
  S_n = |c_1|^2 S_n^{(1)} + |c_2|^2 S_n^{(2)} 
    + S_{n,{\rm mix}},
\label{eq:EE-sep}
\end{equation}
where $S_n^{(1,2)}$ are the entanglement entropies obtained if the state 
were $\ket{\psi_{1,2}}$, and
\begin{equation}
  S_{n,{\rm mix}} = - |c_1|^2 \ln |c_1|^2 - |c_2|^2 \ln |c_2|^2,
\label{eq:S_mix}
\end{equation}
is the entropy of mixing (classical Shannon entropy), suppressed by 
factors of $O({\cal A}_*)$ compared with $S^{(1,2)}_n$.%
\footnote{In the present model, this term is absent for $n < 
 \Delta_{1,2}$.  This is an artifact of the specific toy model 
 adopted here, arising from the fact that two universes cannot 
 be discriminated unless $n$ is larger than one of $\Delta_{1,2}$; 
 see Eq.~(\ref{eq:S_EE-1}).}
This indicates that unless $|c_1|$ or $|c_2|$ is exponentially 
small in $N$, the state $\ket{\psi}$ admits the usual interpretation 
of a superposition of macroscopically different universes with 
$w_{1,2}$ corresponding to $\Delta_{1,2}$.

Similarly, unless a superposition involves exponentially many microstates, 
we find
\begin{equation}
  \ket{\psi} = \sum_i c_i \ket{\psi_i} 
\quad\Rightarrow\quad
  \begin{array}{l}
    \rho_n  = \sum_i |c_i|^2 \rho_n^{(i)},\\
    S_n = \sum_i |c_i|^2 S_n^{(i)}+ S_{n,{\rm mix}},
  \end{array}
\label{eq:rho_n-gen}
\end{equation}
for $n < N/2$, up to exponentially suppressed corrections.  Here, 
$S_{n,{\rm mix}} = -\sum_i |c_i|^2 \ln |c_i|^2$ and is suppressed 
by a factor of $O({\cal A}_*)$ compared with the first term in 
$S_n$.  We thus conclude that the standard many worlds interpretation 
applies to classical spacetimes under any reasonable measurements 
(only) in the limit that $e^{-N}$ is regarded as zero, i.e.\ unless 
a superposition involves exponentially many terms or an exponentially 
small coefficient.  This picture is consonant with the observation 
that classical spacetime has an intrinsically thermodynamic 
nature~\cite{Jacobson:1995ab}, supporting the idea that it 
consists of a large number of degrees of freedom.

We expect that the picture given here persists in the existence of 
excitations on semiclassical backgrounds.  These excitations can be 
represented by non-linear/state-dependent operators at the microscopic 
level, along the lines of Ref.~\cite{Papadodimas:2015jra}.  (For earlier 
work, see Refs.~\cite{Papadodimas:2012aq,Verlinde:2012cy,Nomura:2012ex}.) 
In fact, since entropies associated with the excitations are typically 
subdominant in ${\cal A}_*$~\cite{'tHooft:1993gx,Nomura:2013lia}, 
they have only minor effects on the overall picture.  Therefore, 
we effectively obtain a direct sum structure~\cite{Nomura:2011rb} 
for the Hilbert space
\begin{equation}
  {\cal H}_{B = {\rm FRW},{\cal A}_*}
  \approx \bigoplus_w {\cal H}_w^{{\cal A}_*},
\label{eq:H_prod}
\end{equation}
despite the much more intricate structure of the fundamental Hilbert 
space.

Finally, this fundamental Hilbert space structure suggests that the 
time evolution operator leading to the change of the leaf area is 
also non-linear at the fundamental level.  This does not require the 
time evolution of semiclassical degrees of freedom to be non-linear, 
since the definition of these degrees of freedom would also be 
non-linear at the fundamental level.  Detailed discussions on the 
time evolution in the holographic theory of cosmological spacetimes 
will be presented in Ref.~\cite{NSSW}.

\section*{Acknowledgments}

N.S., F.S., and S.J.W. thank Kavli Institute for the Physics and 
Mathematics of the Universe for hospitality.  This work was supported 
in part by the Department of Energy (DOE), Office of Science, Office 
of High Energy Physics, under contract No.\ DE-AC02-05CH11231, by 
the National Science Foundation under grants PHY-1521446, and by MEXT 
KAKENHI Grant Number 15H05895.  N.S. was supported in part by the 
Simons Heising Physics Fellowship Fund.  F.S. was supported in part 
by the DOE NNSA Stewardship Science Graduate Fellowship.

\end{document}